\documentclass[prb,preprint,unsortedaddress]{revtex4-1} 

\usepackage{amsmath}  
\usepackage{amsfonts} 
\usepackage{graphicx} 
\usepackage{epstopdf}
\graphicspath{{./Graphs/}}
\usepackage{amsmath}
\usepackage{mathtools}
\usepackage{setspace}
\usepackage{hyperref}

\begin{document}

\title{Radial Forcing and Edgar Allan Poe's Lengthening Pendulum}

\author{Matthew McMillan}
\affiliation{Department of Physics, Wheaton College, 501 College Avenue, Wheaton, IL, 60187}

\author{David Blasing}
\affiliation{Department of Physics, Purdue University, 525 Northwestern Avenue, West Lafayette, IN 47907-2036}

\author{Heather M. Whitney}
\email{heather.whitney@wheaton.edu}
\affiliation{Department of Physics, Wheaton College, 501 College Avenue, Wheaton, IL, 60187}


\date{\today}

Copyright 2013 American Association of Physics Teachers. This article may be downloaded for personal use only. Any other use requires prior permission of the author and the American Association of Physics Teachers. \\ \\ 
The following article appeared in Am. J. Phys. 81, 682 (2013) and may be found at \href{http://link.aip.org/link/?ajp/81/682}{McMillan, Blasing, and Whitney, Am. J. Phys. 81, 682 (2013)}

\begin{abstract}
		Inspired by Edgar Allan Poe's \emph{The Pit and the Pendulum}, we investigate a radially driven, lengthening pendulum. We first show that increasing the length of an undriven pendulum at a uniform rate does not amplify the oscillations in a manner consistent with the behavior of the scythe in Poe's story. We discuss parametric amplification and the transfer of energy (through the parameter of the pendulum's length) to the oscillating part of the system. In this manner radial driving may easily and intuitively be understood, and the fundamental concept applied in many other areas. We propose and show by a numerical model that appropriately timed radial forcing can increase the oscillation amplitude in a manner consistent with Poe's story. Our analysis contributes a computational exploration of the complex harmonic motion that can result from radially driving a pendulum, and sheds light on a mechanism by which oscillations can be amplified parametrically. These insights should prove especially valuable in the undergraduate physics classroom, where investigations into pendulums and oscillations are commonplace.
\end{abstract}

\maketitle 

	\section{Introduction}
	
	Physics and literature impart different but often complementary pictures of our world. The disciplines can motivate deeper understandings of each other, and here we use the tools of physics to explore a particular work of literature. \emph{The Pit and the Pendulum}, by Edgar Allan Poe, is a short story that lends itself to an enlightening physical analysis.
	
	Poe published \emph{The Pit and the Pendulum} in 1842.\cite{poe} The grisly story describes an unnamed character on trial during the Spanish Inquisition. The protagonist is convicted and sentenced to a torturous death. Bound to a wooden table, the prisoner gazes upward as a large scythe descends from the ceiling, swinging back and forth. Eventually, it will lower enough to strike and kill the prisoner. Poe explores the swinging pendulum's effects on the prisoner's psyche; here we describe its physics.
	
	Pendulums play a central role in helping physics educators convey fundamental and advanced physics concepts to their students. Analyzing a simple swinging object can require the depths of chaos theory and advanced techniques of differential equations, and has absorbed the ruminations of the likes of Galileo and Newton. On the other hand, one may reduce the analysis to a ``simple'' harmonic oscillator without losing too much insight. Intermediate mechanics texts frequently treat the slightly more advanced harmonic motion of two pendulums attached together, or the path of a pendulum travelling through a viscous fluid. Additional complexities often require a switch to computational modeling, which is sometimes not pursued in a standard undergraduate mechanics course. The complexity of the scythe's behavior in Poe's story behooves a computational approach.\cite{Borrelli}
	
	One might suppose that little new can be said about pendulums. On the other hand, their ubiquity might justify an exhaustive study. We would like to emphasize in this paper an intuitive perspective, which to our knowledge is rarely discussed in the literature. We first discuss the literature on parametric oscillation and pendulums. This motivates a digression on the transfer of energy through the parameter to the system. We consider this perspective an important improvement upon those past, and discuss our theoretical/computational model for the scenario. This model accurately produces the trajectory for uniform lengthening, in close agreement with previous authors. We extend their work by showing that varying the parameter of the length can dramatically amplify the oscillations. Finally, it is noted that this method can properly account for Poe's description, and the insights may fruitfully be applied to other scenarios.
	
	\section{Parametric Amplification, and a Digression on the Transfer of Energy}
	
	We are by no means the first to discuss methods of amplifying a swinging pendulum. The more specific question of applying a radial forcing, or driving, function (i.e., modifying the \textit{length} of the pendulum's string), is not so commonly discussed. A search of the literature turns up just a handful of articles. They broadly fall into one of two categories. One focuses more on resonance phenomena, treating the radially driven pendulum as a parametric oscillator. The other considers energy conservation and the work involved in changing the pendulum's length. To place our work in context, we briefly summarize each.
	
	A paper by Joseph Burns\cite{Burns} and one by Fabrizio Pinto\cite{Pinto} are examples in the first category. Burns considers the equations of motion, treats the pendulum as a parametric oscillator, derives a Hill equation (or the simpler Mathieu equation), and interprets the solution. Pinto takes detailed observations of an elaborate experiment, develops a model based on the ideas of parametric oscillation and resonance (also including Mathieu equations), and shows how his experiment confirms the model. Experiments like Pinto's and our  MATLAB model are explicitly accessible to undergraduate students. The direction of our paper is closer to the second category, so we turn there.
	
	Peter Tea and Harold Falk \cite{Tea} wrote a brief note deriving the quantity of work put into a swing as the swinger periodically raises and lowers his or her center of gravity. Although they approach the question from the perspective of total system energy, they do not go into much detail concerning the general principle we discuss here. The most elaborate contribution from this perspective can be found in the article by Stephen Curry on how children swing by ``pumping,'' or periodically standing and squatting on a swing.\cite{Curry} This effectively varies the radial distance between the pivot and the center of mass. Curry considers the energy added in each period, and derives an exponential amplification. But then he describes the system as a parametric oscillator. He likens it in one case to an LC oscillator where the inductance is varied. In another case he compares it to a nonlinear crystalline solid, through which one higher intensity laser passes through. The laser's intensity alters the parameter of electric susceptibility in just the right way that a half-frequency beam is amplified.
	
	In many cases the perspective of a resonant oscillator, amplified by periodic modulation of a key parameter, is a very helpful way to understand such systems. As in most cases of resonance, usually one takes \textit{frequency} as the important physical quantity. The spectra of such systems includes the harmonic multiples of a fundamental frequency. The degree to which driving affects the amplitude can be measured in terms of how well the driving frequency matches a harmonic of the system.
	
	But we find that sometimes this perspective neglects an important point: the modulated parameter must be connected to the oscillations in just the right way for amplification to occur. For example, the equation of motion for the pendulum is independent of the mass of the bob (indeed, the mass does not even arise in our derivation). We could (in theory) vary the mass, as a parameter, without affecting the oscillations. So in a general context, it is vital that for parametric amplification to succeed, the parameter being altered must be relevant in just the right way.
	
	Looking at energy helps to investigate the connection between parameter and system. In resonant amplification, the system energy increases. Presuming everything is periodic, we can infer that the effect of modulating the parameter is to add a small bit of energy in each cycle. It follows that the parameter should be linked to the relevant, oscillating degrees of freedom in such a way that over each cycle a small amount of energy is added by its modulation. In this way the energy flows from the driving mechanism, through the parameter, and into the oscillating coordinates.
	
	One might be tempted to say that there is a link when the parameter is one of the coordinates of the system (for example, the radius of the pendulum's circular motion). But we must be careful here, for this is not always the case. Consider two examples, first a ball oscillating as it rolls up and down the ``U'' cylinder of a skate-park. The ``z'' axis (considering the ``U'' as a semicircle and using cylindrical coordinates) is a degree of freedom. However, periodically hitting the ball in the $\pm\, \hat{z}$ direction will not affect the oscillations up and down the slope of the U (in the $\hat{\theta}$ direction). The z dimension is \textit{independent}, or ``disconnected,'' from the oscillations. As a second example, consider an LC oscillator, parametrically amplified by altering the inductance. Here the parameter is clearly not a coordinate of the system (current or voltage), but amplification may nonetheless be achieved.
		
	The ``link'' between the parameter and the oscillations can be found in the equations of motion. The parameter could be one of the system coordinates. If it is an oscillating coordinate (e.g., the displacement angle in a pendulum), then we tend to describe it as a \textit{driven harmonic} oscillator, and not a \text{parametric} oscillator. If the parameter is a non-oscillating coordinate, then for successful amplification we expect a ``crossing of terms'' in the equations of motion. The equations of motion should couple the ``parameter'' coordinate (radius) to the oscillating coordinate (angle). Otherwise the energy (or state) of the oscillating coordinate would not change as the parameter is modulated. If the parameter is not an explicit degree of freedom, but another parameter (such as inductance in an LC oscillator), then we should likewise find that parameter in the equations governing the system's oscillations. Not just any parameter will do.
	
	We leave it to the inquisitive reader (perhaps as a class project) to find explicit relations governing the transfer of energy through a general parameter. It might be possible, for example, to give criteria regarding an arbitrary parameter and the Hamiltonian of the system. That is not the purpose of our paper, so we return to our discussion of a lengthening, radially driven pendulum. We do suggest, however, that this be kept in mind when looking at any oscillating system from the perspective of parametric amplification and resonance.
	
	\section{Uniform Lengthening}
	
	The lengthening pendulum problem has been solved analytically and numerically when the lengthening is slow and constant (i.e., \textit{adiabatic}). Such pendulums significantly decrease in angular amplitude as they descend. \cite{Borrelli, Brearley, simoson10, kavanaugh05, Littlewood, Ross, Werner} We present an alternative model, which does not assume a monotonically increasing pendulum length. Instead, we only require that the length function, averaged over one cycle, changes slowly with respect to the pendulum's oscillations. This model may be thought of as a lengthening pendulum with a superimposed parametric amplification by applying radial impulses in phase with the oscillations.
	
\begin{figure}[hbtp]
	\centering
	\includegraphics[width=7cm]{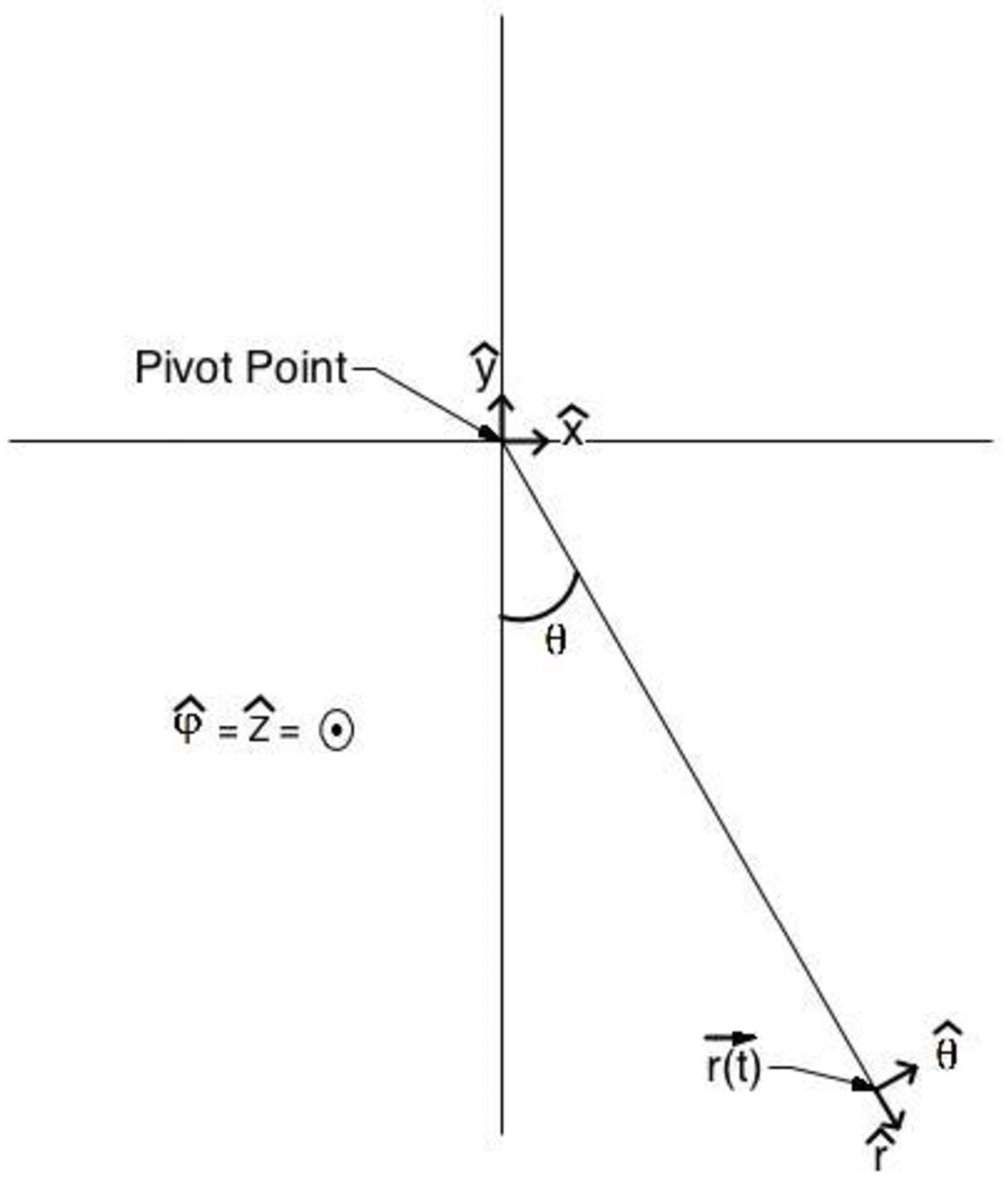}
	\caption{Coordinate systems showing both the inertial $x,y,z$ coordinate system and the non-inertial $r, \theta, \phi$ coordinate system with origin at the pendulum's tip.}
	\label{fig:CoordSys}
\end{figure}

	First we briefly describe a slowly, monotonically lengthening pendulum, similar to that of Simoson as well as Kavanaugh and Moe.\cite{simoson10, kavanaugh05} In particular, this case serves as a test of our computational model, verifying its credibility. The differential equation describing the motion can be derived through Newton's Second Law or via Lagrangian mechanics.
	Figure \ref{fig:CoordSys} shows our coordinate system. 
	
A large scythe lies at the end of a rope with position denoted $\vec{r}(t)$. We assume that the rope never flexes and its mass is negligible compared to the scythe's. The coordinate system is set so the oscillations are in the $xy-$plane. Expressing $\hat{r}$ and $\hat{\theta}$ in terms of $\hat{x}$ and $\hat{y}$ and differentiating with respect to time gives $\dot{\hat{\theta}}= \-\dot{\theta} \hat{r}$ and $\dot{\hat{r}} = \dot{\theta} \hat{\theta}$.
		
	To obtain the acceleration, $\vec{a}$, we take two time derivatives of $\vec{r}(t)=r(t)\hat{r}$. The first yields:
	\begin{equation} \dot{\vec{r}}(t)= \dot{r}(t) \hat{r} + r(t) \dot{\hat{r}} = \dot{r}(t) \hat{r} + r(t) \dot{\theta} \hat{\theta}\label{eq:prepEq1_e} \end{equation}
	
	and the second derivative, along with the triple product rule, gives:
	
	\begin{equation} \ddot{\vec{r}}(t) = [\ddot{r}(t) - r(t) \dot{\theta}^2 ] \hat{r} + [2\dot{r}(t) \dot{\theta} + r(t) \ddot{\theta}] \hat{\theta}. \label{eq:prepEq1_f} \end{equation}
	
	For the present, let the rope's tension force in the $-\hat{r}$ direction cause no radial acceleration (constant lengthening).  The acceleration from gravity is in the $-\hat{y}$ direction, and the $\hat{\theta}$ component is $-g \sin(\theta)$. Equating this with the $\hat{\theta}$ component of Equation \ref{eq:prepEq1_f} yields the differential equation describing the path of a lengthening pendulum:
	
	\begin{equation} r(t) \ddot{\theta} + 2\dot{r}(t) \dot{\theta} + g \sin(\theta) = 0. \label{eq:PendDiffEQ} \end{equation}
	
	Note the familiar form for a harmonic oscillator, $ \ddot{\theta} + \frac{g}{l}\theta = 0 $, when $r(t)$ is a constant \emph{l} and the small angle approximation $\sin(\theta) \approx \theta$ is applied. Equation \ref{eq:PendDiffEQ} can also be derived via Lagrangian mechanics. It is worth noting here the crossing of terms between angular and radial coordinates. The mixing of $r$ and $\theta$ in the unit vector derivatives gives an equation of motion with these coordinates inseparably mixed; thus we might naturally expect the possibility of parametric amplification via one of the coordinates (given oscillation in the other).
	
	The analytical solution of Eq. \ref{eq:PendDiffEQ} involves a superposition of Bessel functions; the reader is directed to Kavanaugh and Moe for more information concerning this approach.\cite{kavanaugh05} Here we employ MATLAB to explore the solutions numerically. The MATLAB function ode45 uses a 4\textsuperscript{th} and 5\textsuperscript{th} order Runge-Kutta integration method to solve systems of first order differential equations. Equation \ref{eq:PendDiffEQ} may be separated into the system below where $ p_1 = \theta $, $ p_2 = \dot{\theta} $, and $ p_3 = r $:
	
	\begin{equation}
	\begin{dcases}
	\dot{p}_1 = p_2(t)  \\
	\dot{p}_2 = \frac{-2\dot{p_3}(t)p_2(t) - g\sin(p_1(t))}{p_3(t)}  \\
	\dot{p}_3 = \text{\emph{constant}}. 
	\end{dcases}
	\label{eq:EqA} \end{equation}
	
	Our MATLAB model gives the numerical solution for $ p_1 $, $ p_2 $, and $ p_3. $
	The initial conditions were $ \theta_0(0)=1.0~\mathrm{rad}, \dot{\theta}_0(0)=0.0~\mathrm( \frac{\text{rad}}{\text{s}}), \text{ and } r(0)=10.0~\mathrm{m}. $

\begin{figure}[hbtp]
	\centering
	\includegraphics[width=12cm]{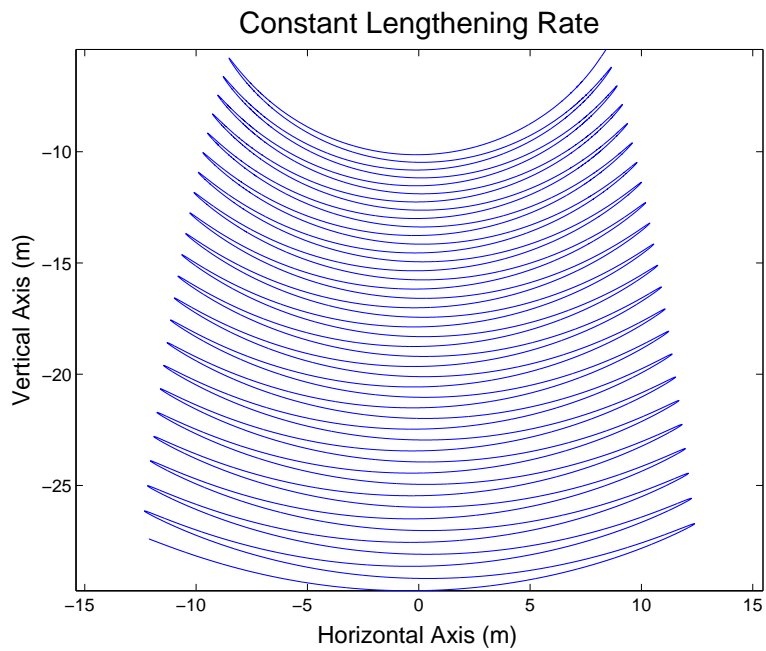}
	\caption{Trajectory of a lengthening pendulum where $r(t) = 0.1\cdot t + 10.0~\mathrm{m}$. Note both the lengthening and the decreasing angular amplitude.}
	\label{fig:NonFoucault}
\end{figure}

\begin{figure}[hbtp]
	\centering
	\includegraphics[width=12cm]{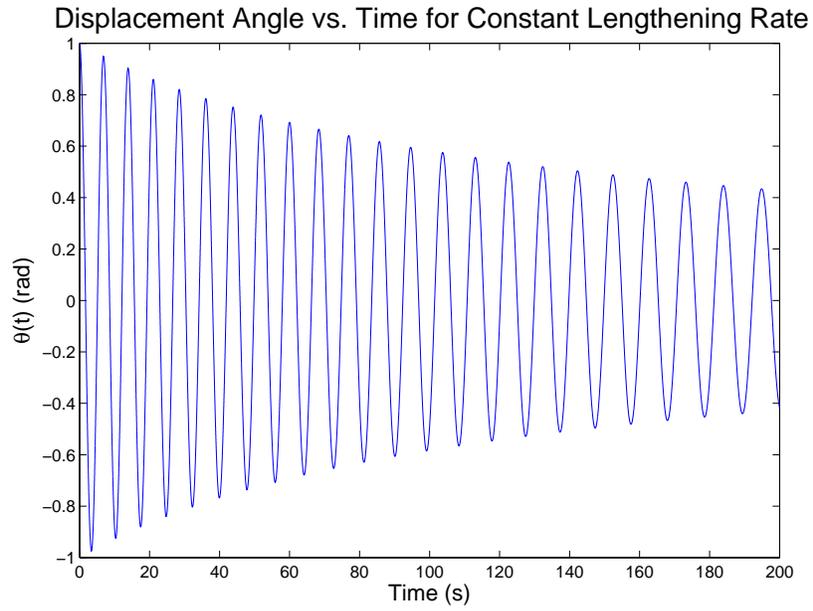}
	\caption{Displacement angle vs. time for the trajectory depicted in Fig.  \ref{fig:NonFoucault}.}
	\label{fig:AngleConstantL}
\end{figure}

	Figure \ref{fig:NonFoucault} plots the trajectory of a regular lengthening pendulum where $r(t) = 0.1\cdot t + 10.0~\mathrm{m},$  $t=0$ to $200~\mathrm{s}$, and with a step size of $1~\mathrm{ms}$. The angular amplitude of this trajectory is plotted in Fig. \ref{fig:AngleConstantL}, and clearly decreases in time. This happens even as the horizontal amplitude increases slightly. The amount of increase of amplitude in Fig. \ref{fig:NonFoucault} is inconsistent with that depicted in \emph{The Pit and the Pendulum}, which Poe describes as dramatic. Poe does not provide specific initial conditions, so we cannot meticulously simulate the trajectory. In any case, a dramatic increase in amplitude cannot be achieved with a uniformly lengthening pendulum. The prisoner says that the descent takes multiple days and that the ceiling is between thirty and forty feet high. An initial length of $\approx 1~\mathrm{m}$, a height $\approx 10~\mathrm{m}$, and the descent time between 2 and 3 days, yields a lengthening rate around $ 5\cdot 10^{-5}~\mathrm{m/s}. $ Such slow lengthening and long duration makes it computationally difficult to simulate and visualize the actual scenario. Accordingly, in these figures, we use a faster lengthening rate and shorter duration (as recorded above) to glean the physical understanding.

\begin{figure}[hbtp]
	\centering
	\includegraphics[width=12cm]{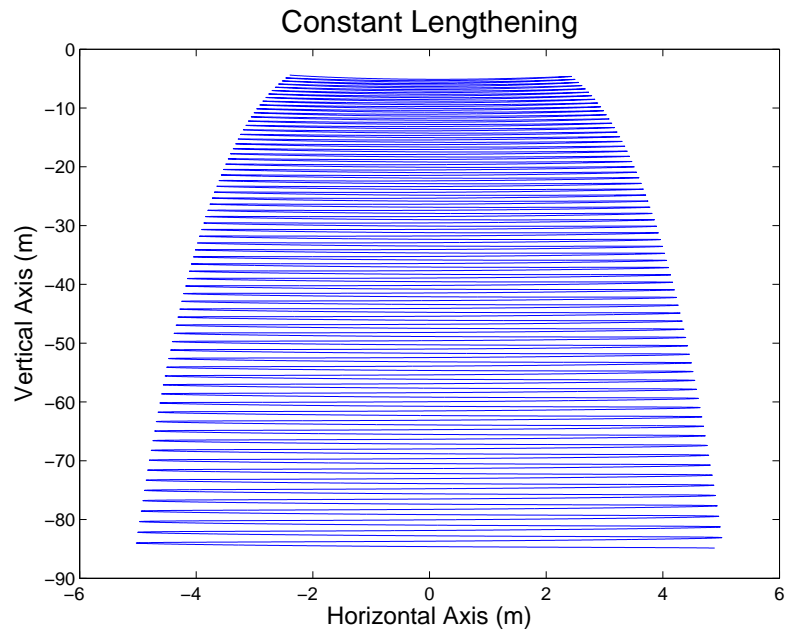}
	\caption{Trajectory of a lengthening pendulum over an extended period.}
	\label{fig:ExtendedPeriodConsLen}
\end{figure}

	To be truer to the story, Fig. \ref{fig:ExtendedPeriodConsLen} shows the path of a lengthening pendulum for an extended period. The parameters and initial conditions are the same as for Fig. \ref{fig:NonFoucault}; the longer time duration clarifies the macro-trajectory but obscures the oscillatory nature. Note again that the angular amplitude decreases while the horizontal amplitude increases. In the next section we present an alternative, physically intuitive explanation, and a simulation with results more consistent with Poe's description.

	\section{Radial Forcing and the Lengthening Pendulum}
	
	It is well understood that applying carefully timed horizontal (or angular) driving forces can increase the oscillation amplitude. Plentiful examples are available in the literature and in many classical mechanics texts. However, we explore here the effects of a \textit{radially} driven pendulum, whose study is far less frequent.
	
	In the context of our digression above, note that the radial coordinate does \textit{not} oscillate in cylindrical coordinates. Nonetheless, it does appear in the angular equation of motion. In cylindrical coordinates, the unit vectors and their derivatives mix. So varying one coordinate (thinking of it as a ``parameter'') can affect a different coordinate. Cartesian coordinates are fully separate, so this never happens. Each Cartesian coordinate has a well defined energy, and this energy stays in that coordinate. This is not so with the pendulum in cylindrical coordinates. The relations between time derivatives of the coordinates themselves allow a carefully timed driving function to transfer energy from the radial into the angular coordinate.
	
\begin{figure}[hbtp]
	\centering
	\includegraphics[width=10cm]{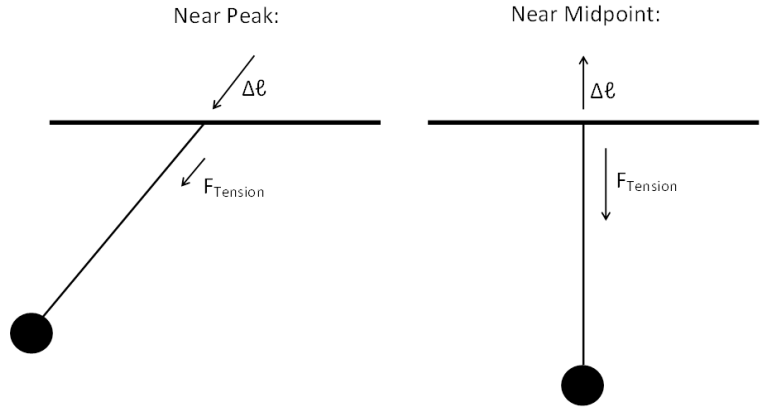}
	\caption{Depiction of the pendulum being pulled in at the bottom and let out at the high point of its cycle.}
	\label{fig:ForcingDiagram}
\end{figure}

	To proceed, we choose a radial driving function. Rather than explicitly work out an analytical solution (which does not exist for an arbitrary driving function and large angles), we develop a hypothesis based on physical intuition, and test it numerically. Our hypothesis runs as follows:

	Suppose now that the pendulum can be pulled in and out through a hole fixed in the ceiling. Let this motion change only the radial position. It has no (immediate) effect on the angular position. At time $t$, let the length be slightly altered by pulling in a small amount $ \varDelta\ell $ during a short time (see Fig. \ref{fig:ForcingDiagram}). Let the radial force on the driving mechanism at $t$ be labelled $F_t$. The work required to make this change is $ F_t\cdot\varDelta\ell $, and is added to the scythe's energy. Furthermore, let the scythe down by the same small $ \varDelta\ell $ at a later time $t'$. This time the scythe does work $ F_{t'}\cdot\varDelta\ell $ on the driving mechanism. The net change in the scythe's energy after both impulses is $ (F_t-F_{t'})\cdot\varDelta\ell $.

	Thus, pulling in when the force is greater and then letting out when the force is lesser increases the scythe's energy. The reverse process, letting the scythe out when in a region of high radial force and in when in a region of low radial force decreases the scythe's energy. So, applying small \textit{radial} impulses can control the scythe's \textit{angular} energy (and thus amplitude).

	The radial force for the swinging pendulum is highest at the bottom and lowest at the peaks of the swing. Additional force is required at the bottom both to maintain the circular motion (the scythe is travelling faster) and to balance gravity. So pulling in near the bottom of the cycle and letting out near the edges will increase the scythe's energy and angular amplitude. Letting out at the bottom and pulling in near the edges will decrease the energy.

	If we let the string out slightly more than we pull in, the average length increases. Thus we arrive at a radially driven, lengthening pendulum. In order to make sure that the amplitude does increase, we need to be careful that there is still a net energy gain. The energy loss for letting more than pulling must not offset the energy gain from letting at the edges and pulling at the bottoms of the cycles.

\begin{figure}[hbtp]
	\centering
	\includegraphics[width=14cm]{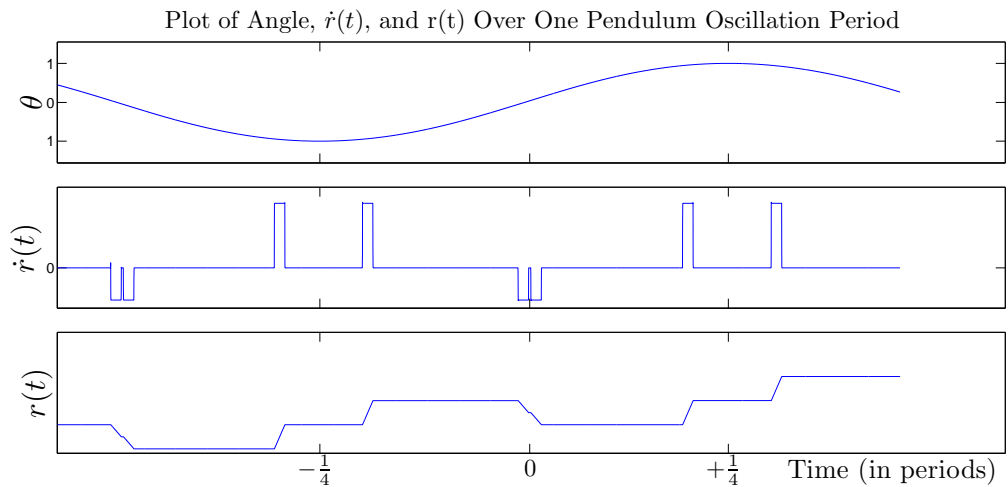}
	\caption{Plot of pendulum state variables over one oscillation period. The positions at which pulling-in and letting-out impulses are applied are displayed.}
	\label{fig:DemoSim}
\end{figure}

	We modeled such a system in MATLAB. Our motivation was observing an amplitude increase in a proof of principle physical experiment that applied radial impulses to a mass on a string. The code models radial impulses by changing the length for short periods and at constant rates. Note that an arbitrary function can in principle be built as a superposition of such functions. The impulse duration and rate were specifiable. Figure \ref{fig:DemoSim} shows the displacement angle $\theta$, $r(t)$, and $\dot{r}(t)$, over one period. This plot helps visualize what happens to the rope length as the pendulum swings. We need a consistent way to specify where in any given cycle is the scythe. It must take into account the varying length, period, and amplitude, while ignoring the particular side the pendulum is on (to preserve the mirror symmetry). We choose a simple notational convention in which the position is specified by a number in the range $[0,1]$; $0$ represents the scythe at the bottom, and $1$ the scythe at either high point of the swing.

\begin{figure}[hbtp]
	\centering
	\includegraphics[width=12cm]{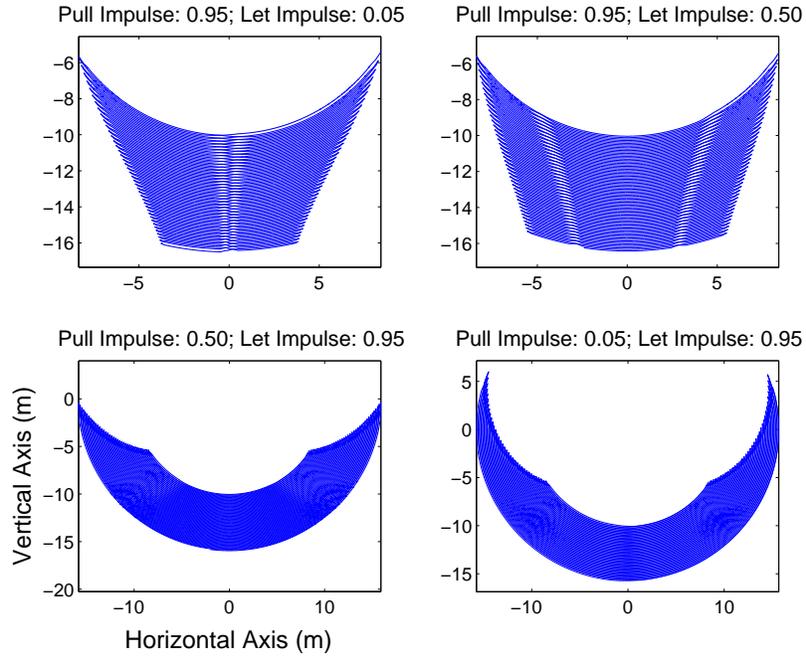}
	\caption{Comparison of the trajectories for different pulling and letting times; this demonstrates that pulling at the bottom (0) and letting at the top (1) can strongly increase the amplitude.}
	\label{fig:RadialImpulses}
\end{figure}

	Figure \ref{fig:RadialImpulses} shows the results of running a simulation over an extended time period. Again, the fractions given for the pulling and letting times represent the fraction of a quarter-oscillation where the impulse is delivered. The letting-out impulses use a speed of $ 0.8~\mathrm{m/s}$, and the pulling-in impulses a speed of $ 0.4~\mathrm{m/s}$. Both last for $ 0.1~ \mathrm{s}$. Thus the average lengthening rate (in all plots) is $ 0.16~ \mathrm{m/cycle}$. The two top plots show that when the pull impulse is higher in the swing than the let impulse, the amplitude decreases faster than in the case of constant lengthening rate. The two bottom plots indicate that when the let impulse is higher than the pull impulse, the amplitude increases. When the scythe is pulled in at 0.05 and let out at 0.95, the amplitude increases dramatically. In this scenario, the scythe can be made to swing all the way around the pivot point by increasing the impulse duration or lengthening rate.

	We should also briefly mention that we apply impulses at 0.05 and 0.95, rather than 0.0 and 1.0 due to the finite duration of the impulses. If we applied an impulse too close to 0.0, it would endure past the low point and interfere with the mirror impulse. The same goes for the impulse at 0.95 rather than 1.0, where an upswing impulse would interfere with a downswing impulse. See Fig. \ref{fig:DemoSim}.

\begin{figure}[hbtp]
	\centering
	\includegraphics[width=12cm]{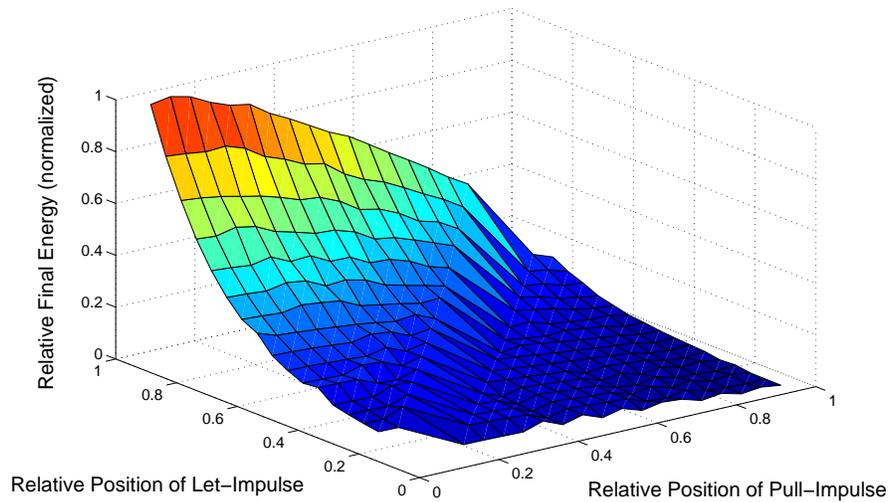}
	\caption{Normalized energy after 300~$\mathrm{s}$ as a function of pulling-in and letting-out impulse positions. Note the strong peak when pulling at the bottom and letting at the high points of each cycle. (Color online.)}
	\label{fig:EndEnergy}
\end{figure}

	Another verification of our hypothesis is obtained by plotting the total energy of the scythe after some fixed length of time $(300~\mathrm{s})$ and varying where in the oscillation the impulses are delivered. This is the content of Fig. \ref{fig:EndEnergy}. In this figure we performed simulations similar to those shown above, and had the pulling-in and letting-out times range over all physical possibilities. Clearly the most energy is added to the scythe when it is pulled up at the bottom and let out at the edge, which is consistent with our hypothesis for how radial impulses increase the scythe's energy. So radial impulses appear to be one mechanism to explain the scythe's behavior in Poe's story.

	\section{Conclusion}
	We reaffirm that a uniformly lengthening pendulum's amplitude does not increase as much as described in \textit{The Pit and the Pendulum}. On the other hand, radial forcing can produce dramatic amplitude increases. Energy flows into the angular oscillations through modulation of the pendulum's length. We gave a theoretical discussion of this process, and verified our hypothesis with a computational model. The results show that this sort of amplification is able to add energy in the dramatic fashion depicted in the story. If the radial forcing were sufficiently small it might go unnoticed, but the amplitude mysteriously would grow. Fortunately for the captive, however, a group of mice eventually chews through the cords tying him down. The prisoner escapes from his descending death with only minor injuries and hopefully a curiosity for the physics behind his experience.

\begin{acknowledgments}

We are grateful to Andrew Morrison for helpful comments and suggestions for this work, as well as to the anonymous reviewers of an earlier draft of this paper for introducing us to the literature on parametric amplification.

\end{acknowledgments}



\section*{Figure captions}

\end{document}